# Electromagnetic coupling between subradiant plasmons and dye molecular excitons analyzed by spectral changes in ultrafast surface-enhanced fluorescence

Tamitake Itoh[1*], Yuko S. Yamamoto[2]

[1]Health and Medical Research Institute, National Institute of Advanced Industrial Science and Technology (AIST), Takamatsu, Kagawa 761-0395, Japan

[2]School of Materials Science, Japan Advanced Institute of Science and Technology (JAIST), Nomi, Ishikawa 923-1292, Japan

*Corresponding author: tamitake-itou@aist.go.jp

**ABSTRACT**

Electromagnetic (EM) coupling between molecular exciton and plasmon has been studied using in Rayleigh scattering or extinction spectroscopy. However, evaluating EM coupling involving subradiant plasmon is challenging because this resonance does not manifest clearly in far-field spectra. In this study, we developed a method to evaluate such coupling using EM enhancement factors ($F_R$) derived from ultrafast surface-enhanced fluorescence (ultrafast SEF). This SEF, which appears as a broad background in surface-enhanced resonant Raman scattering (SERRS) spectra, were measured using silver nanoparticle dimers containing dye molecules within their nanogaps. Our results show that the spectral peaks of $F_R$ for subradiant resonances appear near the dips in Rayleigh scattering spectra. Furthermore, these $F_R$ peaks exhibit blue-shifts during the quenching processes of both ultrafast SEF and SERRS. We examined these static and temporal spectral properties using a coupled oscillator model composed of radiant plasmons, subradiant plasmons, and molecular excitons. The static properties were reproduced by increasing the linewidths of the radiant plasmon resonance, while the temporal properties were captured by decreasing the EM coupling energies between the exciton and both plasmon oscillators. These findings indicate that this methodology is a powerful tool for evaluating EM coupling between subradiant plasmons and molecular excitons.

# I. INTRODUCTION

The large electromagnetic (EM) coupling energy between cavity resonance and exciton resonance is a key to realize the exotic phenomena related to cavity quantum electrodynamics (cavity QED) such as strong-to-ultra strong coupling,[1–3] molecular optomechanics,[4] and polariton chemistry.[5,6] The strong coupling has been investigated using small mode volumes of isolated or aggregated plasmonic nanoparticles.[7–11] Note that the EM coupling energy is proportional to $\sqrt{N_{mol}/V_{HS}}$, where $V_{HS}$ are the mode volume and $N_{mol}$ is the number of molecules inside the mode.[7–11] In particular, the EM coupling energy between a plasmon resonance and dye molecular exciton resonance at nanogaps between silver or gold nanoparticle (NP) dimers reaches several hundred meV due to the extremely small plasmonic mode volume inside the nanogaps.[8–10] Such nanogaps are called hotspots (HSs) and are well known to realize single molecular surface-enhanced resonant Raman scattering (SERRS) spectroscopy.[12,13] The EM enhancement factors originating from this EM coupling at HSs have been clarified to reach to $10^{10}$ for SERRS.[14,15] Such large enhancement factor enabled surface-enhanced hyper Raman spectroscopy even by continuous wave near-infrared laser.[16,17] The large EM coupling energy at HSs exceeds the dephasing energies of both plasmon and exciton resonance, resulting in the hybridized resonance of NP dimer and dye molecules.[8–10] This hybridized resonance largely alters the molecular vibronic states and their dynamics.[5,6] Thus, the relationship between such EM coupling and SERRS has been extensively investigated to apply the SERRS-active strong coupling systems to the platforms for the cavity QED phenomena.[18–20]

The EM coupling at the HSs in dimers has been investigated using dipole-dipole (DD) coupled plasmon, which is a radiant mode, of symmetric small NP dimers with spectral peak change or peak splitting in Rayleigh scattering or extinction of the dimers.[7–10,21,22] However, if the degree of asymmetry or NP size increases, subradiant plasmons such as dipole-quadrupole (DQ) coupled plasmon dominates the EM coupling.[23–25] Such NP systems are more common than symmetric small NP systems for SERRS applications.[20] The subradiant plasmon resonance appears as an unclear dip structure in Rayleigh scattering or extinction spectra.[23–25] Therefore, the observation of EM coupling between DQ-coupled plasmon and molecular exciton is intrinsically difficult with such spectroscopies. We have previously demonstrated that subradiant plasmon resonance can be observed as a spectral peak in ultrafast surface-enhanced fluorescence (SEF) of silver NP or nanowire aggregates containing dye molecules adsorbed in their HSs.[26,27] The ultrafast SEF is fluorescence in which the enhanced rate becomes larger than the vibrational decay rate in the molecular excited state and appears as background emission in the SERRS spectra.[18,28] Thus, the ultrafast SEF is expected to be a useful tool for observing the EM coupling between DQ-coupled plasmons and molecular excitons.[26,27,29]

In this study, we investigated the spectral changes in EM enhancement factors ($F_R$) derived from ultrafast SEF using the HSs of asymmetric silver NP dimers containing several rhodamine 123 (R123) molecules. The DQ-coupled plasmon resonance appeared as a peak and dip in the $F_R$ and Rayleigh scattering spectra, respectively. Both the peak in $F_R$ and dip in Rayleigh scattering exhibited blue shifts during the SERRS quenching process. The static and temporal spectral properties of $F_R$ and Rayleigh scattering were examined using a coupled oscillator model (COM) composed of three oscillators representing a radiant plasmon, a subradiant plasmon, and a molecular exciton. The spectral properties were reproduced well by varying the linewidths of the radiant plasmon resonances and the coupling energies among the oscillators. This reproduction indicates that this method combining ultrafast SEF and COM is useful for evaluating the EM coupling between subradiant plasmons and molecular excitons.

## II. EXPERIMENTAL METHODS

Colloidal silver NPs (mean diameter: ~30 nm, $1.10 \times 10^{-10}$ M) were prepared using the Lee and Meisel method.[30] An equal amounts of the NP dispersion and R123 aqueous solution ($5.0 \times 10^{-8}$ M) were added with NaCl (5 mM) and left for 20 min to form NP aggregates with R123 molecules adsorbed inside the HSs. The final concentrations of the R123 solution ($1.0 \times 10^{-8}$ M) and NP dispersion ($5.5 \times 10^{-11}$ M) were rather higher than the reported single molecule SERRS conditions, as shown by the results of the two-analyte or isotope technique.[31,32] Therefore, SERRS and ultrafast SEF signals may be detected from several dye molecules inside a HS. The sample solution was dropped onto a poly-D-lysine coated glass slide to immobilize the NPs on the slide surface, and the sample plate was covered with another glass plate. The sample plate was then placed on the stage of an inverted optical microscope (IX-71; Olympus, Tokyo, Japan).

The Rayleigh scattering spectra of the single NP dimers were recorded by illuminating white light from a 50-W halogen lamp through a dark-field condenser (numerical aperture, N.A. 0.92). Rayleigh scattering light was collected by an objective (UPlanFL N, Olympus) with N.A. of 0.6, and then the N.A. was increased to 1.3 for collecting the SERRS with ultrafast SEF light. Figures 1(a)–1(c) show images of Rayleigh scattering (N.A. 0.6), extinction (N.A. 1.3), and SERRS with an ultrafast SEF (N.A. 1.3), respectively. Rayleigh scattering and SERRS with ultrafast SEF light from single silver NP dimer located at the centre of the image were selected using a pinhole and sent to a polychromator equipped with a thermoelectrically cooled charge-coupled device assembly (DV 437-OE-MCI, Andor, Japan). Figure 1(d) shows a typical Rayleigh scattering cross-section ($\sigma_{sca}$) spectrum obtained by dividing the Rayleigh scattering intensity spectrum by the spectrum of the white light source to exclude spectral modulation by the light source. The SERRS and ultrafast SEF spectra of the identical dimers were measured by illuminating the sample with an unpolarized blue laser beam (457 nm, 3.5 W/cm$^2$; CW DPSS 457, Laser Create) that has light energy (2.71 eV) higher than that of molecular absorption (~2.42 eV). The advantage of using a higher excitation energy laser is to expand the spectral window for the $F_R(\omega)$ spectra, as indicated in Fig. 1(e).[27] We selected a low-groove-density grating (150 grooves/mm) for the polychromator to merge the detailed structures in the SERRS spectra, obtaining smooth $F_R(\omega)$ spectra, as indicated in Fig. 1(e).

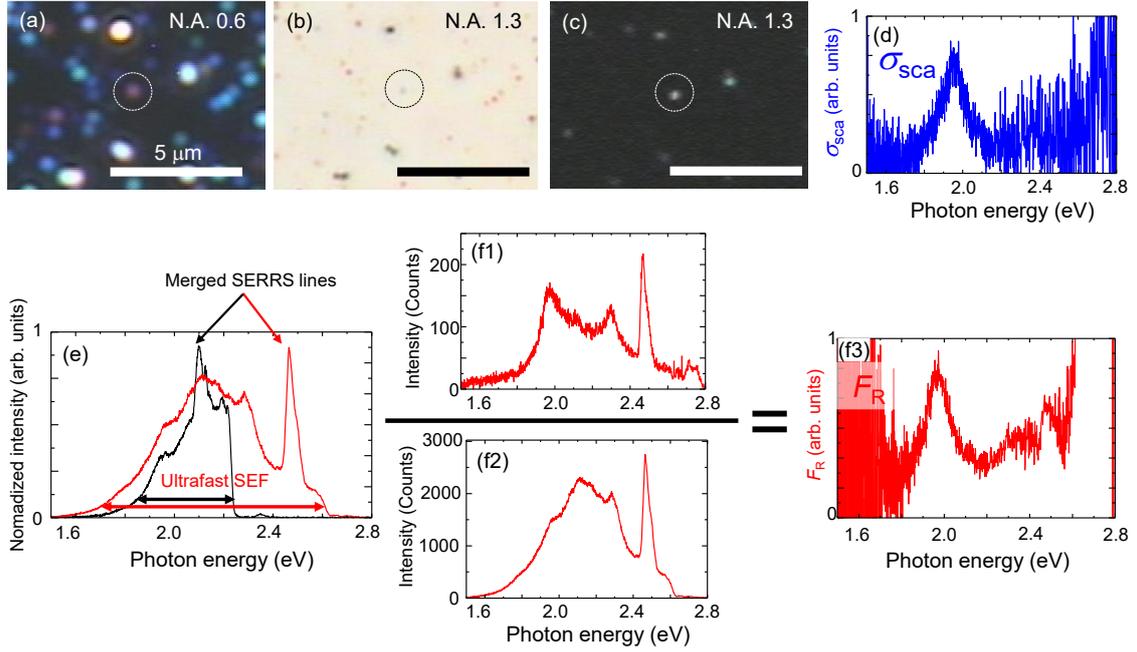

FIG. 1. (a)–(c) Rayleigh scattering, extinction, and ultrafast SEF (including SERRS) images of silver NPs or NP dimers, respectively. The dotted circles indicate detection areas. (d) Typical $\sigma_{sca}(\omega)$ spectrum obtained by dividing the Rayleigh scattering intensity spectrum by the white light source spectrum. (e) Spectra of SERRS with ultrafast SEF with excitation laser energy of 2.33 eV (532 nm) (black curve) and 2.71 eV (457 nm) (red curve). The laser lines were omitted by the notch filters. SERRS lines are merged due to the use of low groove density grating. (f) Procedure for deriving $F_R(\hbar\omega)$. (f1) and (f2) Spectra of SERRS with ultrafast SEF of single silver NP dimer and large silver NP aggregate, respectively. (f3) The obtained spectrum of $F_R(\omega)$.

The procedure used to experimentally obtain the $F_R(\omega)$ spectra is illustrated in Figs. 1(f1)–1(f3). Figures 1(f1) and 1(f2) show the ultrafast SEF spectra of a single NP dimer and large NP aggregate, respectively. The $F_R(\omega)$ spectrum of large NP aggregates is flat because it is the sum of various $F_R(\omega)$ spectra from a large number of HSs.[33] Therefore, the $F_R$ spectrum in Fig. 1(f3) was derived by dividing the ultrafast SEF spectrum in Fig. 1(f1) by that in Fig. 1(f2). The contributions of the radiant and subradiant plasmon resonances to $F_R$ were evaluated by comparing the obtained $F_R(\omega)$ [Fig. 1(f3)] with $\sigma_{sca}(\omega)$ of a single NP dimer [Fig. 1(d)]. In short, the contribution of the subradiant resonance appears as a spectral peak in $F_R(\omega)$ and a dip in $\sigma_{sca}(\omega)$.[26,27,29]

## III. RESULTS AND DISCUSSION

We found two different types of spectral changes in $\sigma_{sca}(\omega)$ and $F_R(\omega)$ for single-NP dimers during SERRS quenching. One is the peak change in $\sigma_{sca}(\omega)$ synchronized with the peak changes in $F_R(\omega)$; dimers exhibiting this type of synchronized spectral changes

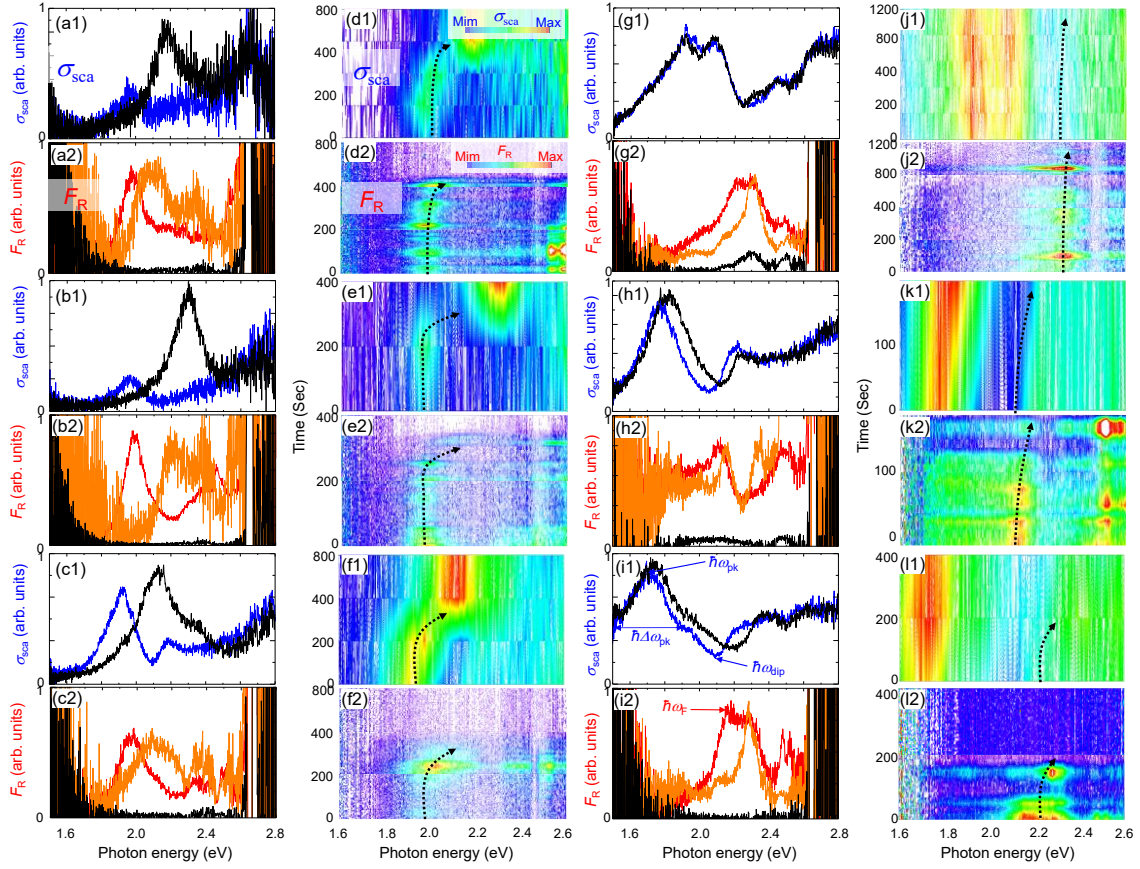

FIG. 2. (a1)–(c1) $\sigma_{sca}(\omega)$ spectra before (blue curves) and after (black curves) SERRS quenching for Type I dimers. (a2)–(c2) $F_R(\omega)$ spectra before (blue curves), just before (orange curves), and after (black curves) SERRS quenching for Type I dimers. (d1)–(f1) $\sigma_{sca}(\omega)$ spectra during SERRS quenching process expressed as contour maps for the Type I dimers of (a1)–(c1). (d2)–(f2) $F_R(\omega)$ spectra during SERRS quenching process expressed as contour maps for the Type I dimers of (a2)–(c2). (g1)–(i1) $\sigma_{sca}(\omega)$ spectra before (blue curves) and after (black curves) SERRS quenching for Type II dimers. (g2)–(i2) $F_R(\omega)$ spectra before (blue curves), just before (orange curves), and after (black curves) SERRS quenching for Type II dimers. (j1)–(l1) $\sigma_{sca}(\omega)$ spectra during SERRS quenching process expressed as contour maps for the Type II dimers of (g1)–(i1). (j2)–(l2) $F_R(\omega)$ spectra during SERRS quenching process expressed as contour maps for the Type II dimers of (g2)–(i2). The dotted arrows in all contour maps indicate the time traces of maximum position changes in $F_R(\omega)$. The definitions of $\hbar\omega_{pk}$, $\hbar\omega_{dip}$, $\hbar\Delta\omega_{pk}$, and $\hbar\omega_F$ are indicated in (i1) and (i2).

are hereafter designated as "Type I". The other is the change in the dips of $\sigma_{sca}(\omega)$ synchronized with the peak changes in $F_R(\omega)$; dimers exhibiting this type of spectral changes are hereafter designated as "Type II", Figures 2(a1) and 2(a2) to Figs. 2(c1) and 2(c2) show typical Type I examples. Their spectral peaks in $\sigma_{sca}(\omega)$ and $F_R(\omega)$ occur at the common positions and commonly exhibit blue shift during SERRS quenching process. This spectral change in $\sigma_{sca}(\omega)$ has been clarified as decrease in EM coupling energy between molecular excitation and radiant plasmon.[10,21,22] These common spectral change between $F_R(\omega)$ and $\sigma_{sca}(\omega)$ indicates that the $F_R(\omega)$ is generated by the radiant plasmon.[10,21] Figures 2(d1) and 2(d2) to Figs. 2(f1) and 2(f2) are spectral changes in $\sigma_{sca}(\omega)$ and $F_R(\omega)$ expressed as contour maps. These peak shifts in $F_R(\omega)$ synchronized with those in $\sigma_{sca}(\omega)$, confirming the generation of $F_R(\omega)$ by radiant plasmons. Figures 2(g1) and 2(g2) to Figs. 2(i1) and 2(i2) show typical Type II examples. The spectral peaks in $F_R(\omega)$ appeared around the spectral dips (not peaks) in $\sigma_{sca}(\omega)$ and these spectral peaks and dips exhibited blue shifts during the SERRS quenching process. The appearance of $F_R(\omega)$ peaks around the dips in $\sigma_{sca}(\omega)$ indicates that the $F_R(\omega)$ is generated by subradiant plasmon,[24–27] because subradiant plasmon resonance generates a dip in $\sigma_{sca}(\omega)$ due to coherent light energy transfer from radiant to subradiant plasmon resonance.[29] Figures 2(j1) and 2(j2) to Figs. 2(l1) and 2(l2) are spectral changes in $\sigma_{sca}(\omega)$ and $F_R(\omega)$ expressed as contour maps. The peak shifts in $F_R(\omega)$ appeared to synchronize with the dip in $\sigma_{sca}(\omega)$, indicating that the decrease in the EM coupling energy between the molecular excitation

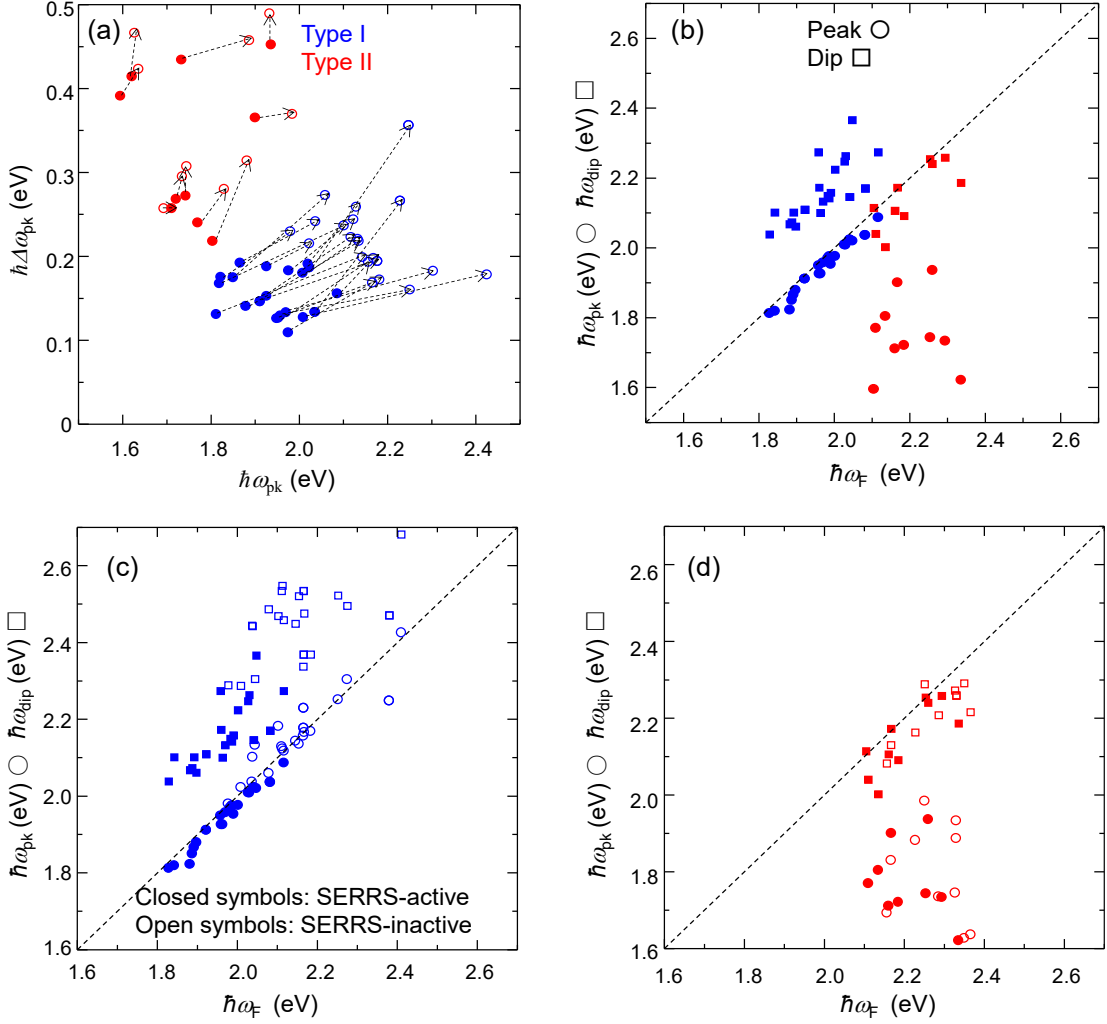

FIG. 3. (a) Relationship between $\hbar\omega_{pk}$ and $\hbar\Delta\omega_{pk}$ for Type I (blue circles) and Type II (red circles) dimers before (closed circles) and after (open circles) quenching SERRS and ultrafast SEF. The arrows indicate the direction of changes for each dimer. (b) Relationship between $\hbar\omega_F$ and $\hbar\omega_{pk}$ (blue circles) or $\hbar\omega_{dip}$ (blue squares) for Type I dimers and relationship between $\hbar\omega_F$ and $\hbar\omega_{pk}$ (red circles) or $\hbar\omega_{dip}$ (red squares) for Type II dimers. (c) Relationship between $\hbar\omega_F$ and $\hbar\omega_{pk}$ (blue circles) or $\hbar\omega_{dip}$ (blue squares) for Type I dimers before (closed symbols) and after (open symbols) quenching SERRS and ultrafast SEF. (d) Relationship between $\hbar\omega_F$ and $\hbar\omega_{pk}$ (red circles) or $\hbar\omega_{dip}$ (red squares) for Type II dimers before (closed symbols) and after (open symbols) quenching SERRS and ultrafast SEF.

and subradiant plasmon facilitates this spectral change. This spectral change is a central issue addressed in this study. Note that there is an intermediate type between Type I and II; however, we excluded such dimers to clarify the analysis (See supplementary material

SI for the intermediate type).

The features of the spectral changes in the $\sigma_{sca}(\omega)$ and $F_R(\omega)$ of the multiple dimers are summarized in Fig. 3. Here, $\hbar\omega_{pk}$, $\hbar\omega_{dip}$, and $\hbar\Delta\omega_{pk}$ denote the peak energy, dip energy, and linewidth of $\sigma_{sca}(\omega)$, respectively, as indicated in Fig. 2(i1); $\hbar\omega_F$ denotes the peak energy of $F_R(\omega)$, as as indicated in Fig. 2(i2). Figure 3(a) shows the relationship between $\hbar\omega_{pk}$ and $\hbar\Delta\omega_{pk}$ of radiant plasmon resonance appearing in $\sigma_{sca}(\omega)$ before and after SERRS quenching for Type I and II dimers. The Type I dimers clearly show the blue shift in peak energy by approximately 200 meV with linewidth broadening. The degree of peak blue shift and broadening of the Type II dimers appear rather minor compared to those of the Type I dimers. One can also notice that the $\hbar\Delta\omega_{pk}$ of Type II dimers are much broader than those of Type I dimers. Figure 3(b) shows the relationship between the $\hbar\omega_F$ and the $\hbar\omega_{pk}$ (or $\hbar\omega_{dip}$) in $\sigma_{sca}(\omega)$ for the Type I and II dimers. The $\hbar\omega_{pk}$ and $\hbar\omega_{dip}$ is indicated by circles and squares, respectively. For the Type I dimers (blue symbols), the $\hbar\omega_F$ matched well with $\hbar\omega_{pk}$, whereas the $\hbar\omega_{dip}$ always appeared at the higher energy side of the $\hbar\omega_F$. In contrast, for the Type II dimers (red symbols), the $\hbar\omega_F$ matched well with the $\hbar\omega_{dip}$ and the $\hbar\omega_{pk}$ always appeared the lower energy side of the $\hbar\omega_F$. Figure 3(c) shows the relationship between the $\hbar\omega_F$ and the $\hbar\omega_{pk}$ (or $\hbar\omega_{dip}$) before and just before SERRS quenching for the Type I dimers. Both $\hbar\omega_{pk}$ and $\hbar\omega_{dip}$ exhibit blue shifts and the $\hbar\omega_{pk}$ matches better with the $\hbar\omega_\text{F}$. Figure 3(d) shows the same relationship as Fig. 3(c) for the Type II dimers. Both $\hbar\omega_\text{pk}$ and $\hbar\omega_\text{dip}$ exhibit blue shifts and the $\hbar\omega_\text{dip}$ matches well with the $\hbar\omega_\text{F}$. The spectral shifts in Fig. 3(c) are larger than those in Fig. 3(d). All the tendencies in Figs. 3(a)–3(d) are consistent with the spectral changes in $\sigma_\text{sca}(\omega)$ and $F_\text{R}(\omega)$ shown in Fig. 2.

We summarize the static and temporal spectral properties observed in $\sigma_\text{sca}(\omega)$ and $F_\text{R}(\omega)$ as follows:

(1) There are two types of relationship between $\sigma_\text{sca}(\omega)$ and $F_\text{R}(\omega)$. Type I: the $\hbar\omega_\text{F}$ is similar to the $\hbar\omega_\text{pk}$, and Type II: the $\hbar\omega_\text{F}$ is similar to the $\hbar\omega_\text{dip}$.

(2) The $\hbar\Delta\omega_\text{pk}$ for Type II dimers is broader than that for Type I dimers. The $\hbar\omega_\text{pk}$ of Type II dimers is rather smaller than that of Type I dimers.

(3) Both $\hbar\omega_\text{pk}$ and $\hbar\omega_\text{dip}$ exhibit blue shift during SERRS quenching processes. The $\hbar\omega_\text{F}$ also exhibits blue shift during these processes.

(4) The amount of blue shift in the $\hbar\omega_\text{pk}$ and $\hbar\omega_\text{F}$ of Type II dimers are smaller than those of Type I dimers.

In our previous studies, we explored the origins of radiant and subradiant plasmons based on the comparison of spectral features of $\sigma_\text{sca}(\omega)$ and $F_\text{R}(\omega)$ with scanning electron microscopy (SEM) images of dimers.[24] The SEM images revealed that Type I and II dimers have symmetric and asymmetric shapes, respectively.[24] The origins of radiant and subradiant plasmons were determined from finite-difference time-domain (FDTD) calculations (EEM-FDM Version 5.1, EEM Co., Ltd., Japan) based on the SEM images.[24] The results indicated that the radiant plasmon is the coupled mode between two dipole plasmons of two NPs (i.e., DD coupled plasmon), whereas the subradiant plasmon is the coupled mode between dipole plasmon of smaller NP and quadrupole plasmon of larger NP (i.e., DQ coupled plasmon).[23,24] The origins of temporal blue shifts in $\hbar\omega_{pk}$ during SERRS quenching processes were determined as the decrease in EM coupling energy between a DD-coupled plasmon and a molecular exciton using COM.[10,21,22] The analysis of ratios between SERRS and SEF intensities revealed the increase in the effective distance between the molecules and metal surfaces at HSs during SERRS quenching.[33] This result supports that the decrease in EM coupling energy was induced by the instability and desorption of the dye molecules inside the HS.[21] Therefore, we evaluated the observed spectral changes in $\sigma_{sca}(\omega)$ and $F_R(\omega)$ [summarized in (1)–(4)] as the changes in EM coupling energy between a plasmon and a molecular exciton using COM. We assumed that the EM coupling energy between the molecular exciton and DD-coupled plasmon is identical to the EM coupling energy between the molecular excitons and DQ-coupled plasmons because both plasmons share a common HS, where the EM coupling energy is proportional to $\sqrt{N_{\text{mol}}/V_{\text{HS}}}$.[21]

The EM coupling between a molecular exciton and a plasmon has been described by Jaynes–Cummings model.[1,2] The Jaynes–Cummings model composed of quantum emitter, dipole (radiant) plasmon, and quadrupole (non-radiant) plasmon was also reported to evaluate the effect of non-radiant plasmon on strong coupling.[34] The COM phenomenologically approximates the Jaynes–Cummings model in the case that coherent coupling is dominant (or quantum fluctuations can be neglected).[1,2] Indeed, the COM well reproduces plasmon-induced transparency and strong coupling between a molecular exciton and a plasmon.[10,11,21,22,35,36] Thus, we evaluated the observed spectral changes in $\sigma_{\text{sca}}(\omega)$ and $F_R(\omega)$ [summarized in (1)–(4)] using the COM composed of three oscillators representing molecular exciton, DD-coupled plasmon, and DQ-coupled plasmon.

The procedure for evaluating the changes in $\sigma_{\text{sca}}(\omega)$ and $F_R(\omega)$ using COM is as follows: First, we calculated the DD- and DQ-coupled plasmons in $\sigma_{\text{sca}}(\omega)$ and $F_R(\omega)$ using the FDTD method by varying the degree of asymmetry of the dimers. Then, we reproduced these FDTD calculations using a COM composed of DD- and DQ-coupled plasmons of dimers to derive their coupling energy and spectral properties. Finally, we added the oscillators representing the molecular exciton of a two-level system to the COM and evaluated the observed spectral features in $\sigma_{\text{sca}}(\omega)$ and $F_R(\omega)$, as summarized in (1)–

(4).

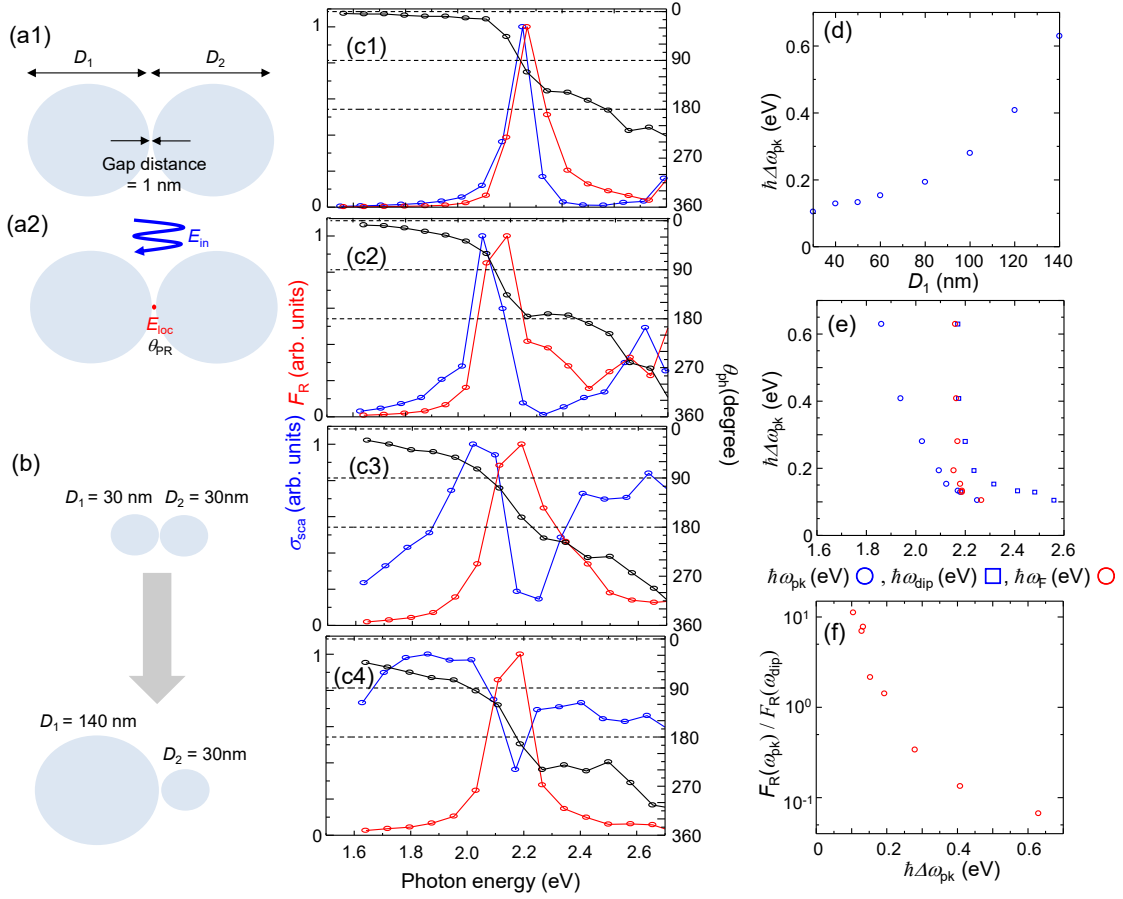

FIG. 4. (a1) Schematics of NP dimer showing $D_1$, $D_2$, and gap distance (1 nm). (a2) Position of $\theta_{ph}$ and $E_{loc}$ indicated by the red dot. (b) Schematics of the increase in the degree of asymmetry of NP dimer. Spectra of $\sigma_{sca}(\omega)$ (blue curve) and $F_R(\omega)$ (red curve) normalized by their maxima, and $\theta_{ph}(\omega)$ (black curve) for dimers with $D_1$ of (c1) 30, (c2) 60, (c3) 100, and (c4) 140 nm; $D_2$ = 30 nm. The upper, middle, and lower dashed lines correspond to $\theta_{ph}$ of 90°, 135° and 180°, respectively. (d) Relationship between $D_1$ and $\hbar\Delta\omega_{pk}$ for dimers with $D_1$ of 30, 60, 100, and 140 nm; $D_2$ = 30 nm. (e) Relationship between $\hbar\Delta\omega_{pk}$ and the three positions: $\hbar\omega_{pk}$ (blue circles), $\hbar\omega_{dip}$ (blue squares), and $\hbar\omega_F$ (red circles). (f) Relationship between $\hbar\Delta\omega_{pk}$ and $F_R(\omega_{pk})/F_R(\omega_{dip})$ (red circles).

Figure 4(a1) illustrates dimers composed of two NPs with diameters $D_1$ and $D_2$ for the FDTD calculation. The complex refractive indices of the silver NPs were obtained from Ref. 38. We derived $F_R(\omega)$ at the gap center of the dimer, where there is a HS, as $|E_{loc}(\omega)|^2/|E_{in}(\omega)|^2$. $|E_{loc}(\omega)|$ and $|E_{in}(\omega)|$ are the local electric field amplitudes inside the HS and the incident electric field amplitude, respectively. The gap distance between the two NPs was sets to be 1.0 nm considering the size of the adsorbed dye molecule in the gap.[24–26] The position of $E_{loc}(\omega)$ and its phase retardation $\theta_{ph}(\omega)$ against $E_{in}(\omega)$ are indicated in Fig. 4(a2). Figure 4(b) schematically shows the increase in the degree of dimer asymmetry. Figures 4(c1)–4(c4) show the spectral changes in $\sigma_{sca}(\omega)$, $F_R(\omega)$, and $\theta_{ph}(\omega)$ with increasing $D_1$ from 30 to 120 nm for fix $D_2 = 30$ nm. The $\theta_{ph}(\omega)$s of 90° and 180° correspond to DD- and DQ-coupled resonance, respectively.[24] As their degree of dimer asymmetry increases, the $\hbar\omega_F$ moves from the $\hbar\omega_{pk}$ to the $\hbar\omega_{dip}$. These shifts indicate that the plasmon resonance generating $F_R(\omega)$ changes from DD- to DQ-coupled resonance.[24–26] Figures 4(d) exhibits the relationship between $D_1$ and $\hbar\Delta\omega_{pk}$. As the $D_1$ increases, the $\hbar\Delta\omega_{pk}$ also increases due to the increase in radiation loss.[37] Figures 4(e) shows the relationship between $\hbar\Delta\omega_{pk}$ and the three positions such as $\hbar\omega_{pk}$, $\hbar\omega_{dip}$, and $\hbar\omega_F$. The $\hbar\omega_F$ jumps from $\hbar\omega_{pk}$ to $\hbar\omega_{dip}$ around the $\hbar\Delta\omega_{pk}$ of 0.1–0.2 eV. This jump occurs via coherent excitation light energy transfer from the DD- to the DQ-coupled resonance. This energy transfer was confirmed by the changes of $\theta_{ph}(\omega_F)$ from 90° to 180°, indicating the values for DD- and DQ-coupled resonance, respectively.[24,26,29] In other words, the electric field energy soured in DD-coupled resonance further resonates with DQ-coupled resonance by near-field interaction and generates $F_R(\omega_F)$ at $\theta_{ph}(\omega_F)$ of 180°.[29,39] Figure 4(f) shows the ratio $F_R(\omega_{pk})/F_R(\omega_{dip})$ as a function of $\hbar\Delta\omega_{pk}$. As the $\hbar\Delta\omega_{pk}$ increases, $F_R(\omega_{dip})$ becomes larger than $F_R(\omega_{pk})$, supporting the above discussion that the plasmon resonance generating $F_R(\omega)$ changes from DD- to DQ-coupled resonance with increasing the degree of dimer asymmetry.

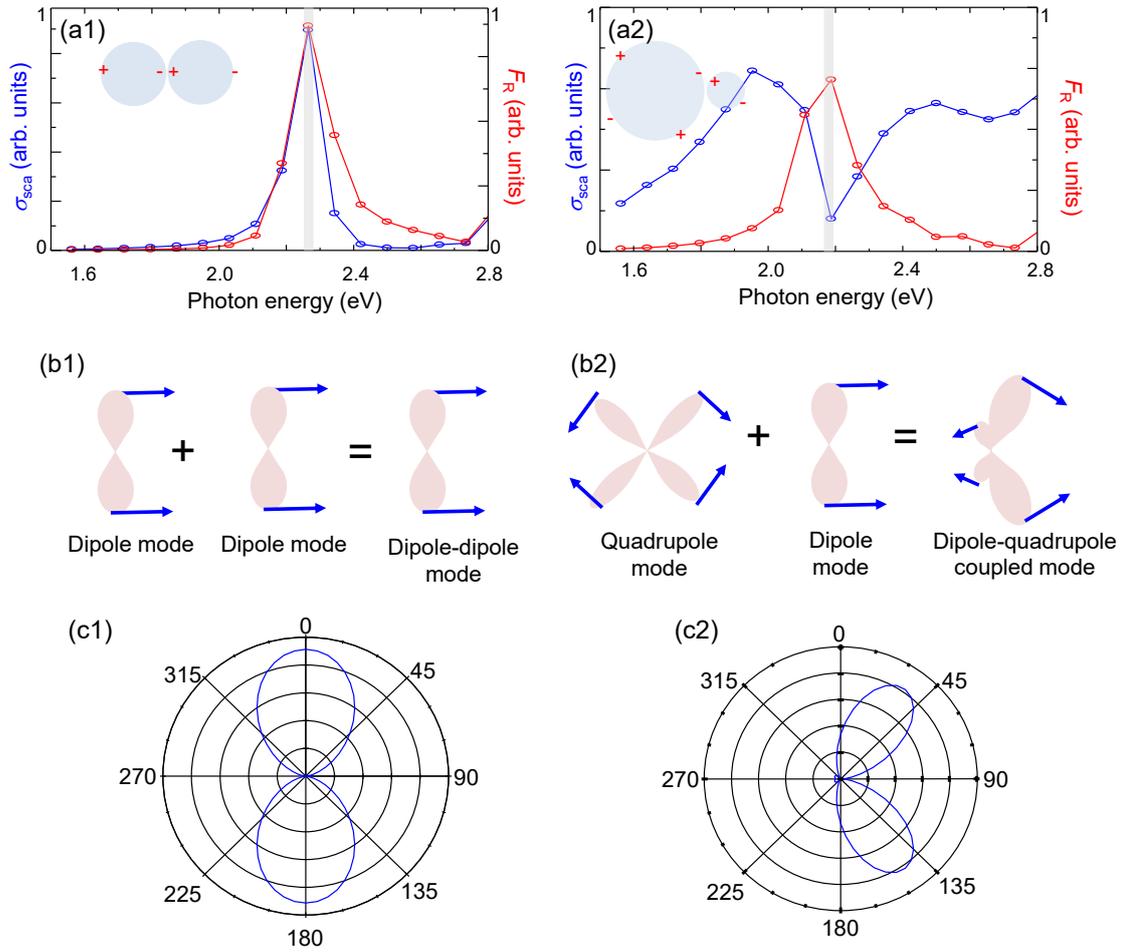

FIG. 5. (a1) and (a2) Normalized spectra of $\sigma_{sca}(\omega)$ (blue curve) and $F_R(\omega)$ (red curve) for symmetric ($D_1 = D_2 = 30$ nm) and asymmetric ($D_1 = 140$ nm; $D_2 = 30$ nm) dimers, respectively. The gap distance is 1 nm. Insets are images of charge distributions. Schematics of radiation patterns for (b1) DD- and (b2) DQ-coupled resonance. The blue arrows indicate the directions of the electric field vectors. Polar diagram of the radiation patterns for Rayleigh scattering light of (c1) DD-coupled plasmon at the gray line in (a1) and (c2) DQ-coupled plasmon at the gray line in (a2).

Furthermore, we analyzed the radiation properties of the DD- and DQ-coupled plasmons of symmetric and asymmetric dimers, which correspond to Type I and II dimers, respectively. Figures 5(a1) and 5(a2) show the spectra of $\sigma_{sca}(\omega)$ and $F_R(\omega)$ for the symmetric and asymmetric dimers, respectively. The two vertical grey lines at $\hbar\omega_{pk}$ and $\hbar\omega_{\text{dip}}$ indicate DD- and DQ-coupled resonance maxima, respectively. The charge distribution of the DD-coupled plasmon in Fig. 5(a1) implies that this dimer exhibits a dipole radiation pattern, because both dipole radiation fields constructively interfere with each other, as shown in Fig 5(b1).[25] The charge distribution of DQ-coupled plasmon is shown in Fig. 5(a2). Figure 5(b2) shows the destructive interference of the radiation field between the quadrupole and the dipole when the radiation is directed toward the larger NP side. Constructive interference also occurs when radiation is directed toward the smaller NP side, as shown in Fig. 5(b2). Such destructive and constructive interference may bend the dipole-like radiation.[25,40] To check the radiation properties, the far-field radiation patterns at the spectral positions of $\hbar\omega_{\text{pk}}$ and $\hbar\omega_{\text{dip}}$ were examined with FDTD calculation. Figures 5(c1) and 5(c2) show the polar diagrams of the radiation patterns at $\hbar\omega_{\text{pk}}$ and $\hbar\omega_{\text{dip}}$ in Figs. 5(a1) and 5(a2), respectively. The radiation pattern in Fig. 5(c1) exhibits dipole features. Figure 5(c2) shows bending to the side of the smaller NP, which is evidence of the contribution of the DQ-coupled plasmon to radiation. The important point of these polar diagrams for analysis using COM is that we should treat the ultrafast SEF as radiation from both DD- and DQ-coupled plasmons.

Here, we explain the derivation of $\sigma_{\text{sca}}(\omega)$ and $F_R(\omega)$ using the COM. The driving force of the COM for $\sigma_{\text{sca}}(\omega)$ is light. In contrast, the driving force for $F_R(\omega)$ is the transition of excited excitons triggered by the surrounding EM vacuum fluctuation inside the HS.[2,11,36]

Thus, $\sigma_{sca}(\omega)$ and $F_R(\omega)$ were independently derived for the analysis.

The equations of motion for the coupled oscillator for $\sigma_{sca}(\omega)$ are

$$\frac{\partial^2 x_m(t)}{\partial t^2} + \gamma_m \frac{\partial x_m(t)}{\partial t} + \omega_m^2 x_m(t) - g_{mDD}\frac{\partial x_{DD}(t)}{\partial t} - g_{mDQ}\frac{\partial x_{DQ}(t)}{\partial t} = 0, \quad (1)$$

$$\frac{\partial^2 x_{DD}(t)}{\partial t^2} + \gamma_{DD}\frac{\partial x_{DD}(t)}{\partial t} + \omega_{DD}^2 x_{DD}(t) + g_{mDD}\frac{\partial x_m(t)}{\partial t} + g_{DDDQ}\frac{\partial x_{DQ}(t)}{\partial t} = P(t), \quad (2)$$

$$\frac{\partial^2 x_{DQ}(t)}{\partial t^2} + \gamma_{DQ}\frac{\partial x_{DQ}(t)}{\partial t} + \omega_{DQ}^2 x_{DQ}(t) + g_{mDQ}\frac{\partial x_m(t)}{\partial t} + g_{DDDQ}\frac{\partial x_{DD}(t)}{\partial t} = 0. \quad (3)$$

where $x_m$, $x_{DD}$, and $x_{DQ}$ are the coordinates of a molecular exciton, DD-, and DQ-coupled plasmon oscillation, respectively; $\gamma_m$, $\gamma_{DD}$, and $\gamma_{DQ}$ are the dephasing rates of the excitonic, the DD-, and the DQ-coupled plasmon oscillation, respectively; $\omega_m$, $\omega_{DD}$, and $\omega_{DQ}$ are the angular resonance frequencies of the exciton, the DD-, and the DQ-coupled plasmon oscillators, respectively; $g_{mDD}$, $g_{mDQ}$, and $g_{DDDQ}$ are the coupling rates between the exciton and DD-coupled plasmon, between the exciton and DQ-coupled plasmon, between the DD- and DQ-coupled plasmon, respectively; and $P(t)$ denotes the driving forces representing incident light. We assumed in Eqs. (1)–(3), the exciton oscillator and the DQ-coupled plasmon oscillator are entirely driven by the DD-coupled plasmon oscillator.[2,10,11] By setting $P(t) = Pe^{-i\omega t}$, where $P$ and $\omega$ are the amplitude and incident light frequency, respectively, the intensities at $x_m$, $x_{DD}$, and $x_{DQ}$ can be derived from Eqs.

(1)–(3) as follows:

$$|x_m(\omega)|^2 = \left|\frac{i(D_{DQ}g_{mDD}\omega - ig_{mDQ}g_{DDDQ}\omega^2)}{D_m D_{DD} D_{DQ} - D_{DQ}g_{mDD}^2\omega^2 - D_{DD}g_{mDQ}^2\omega^2 - D_m g_{DDDQ}^2\omega^2}\right|^2, \quad (4)$$

$$|x_{DD}(\omega)|^2 = \left|\frac{D_m D_{DQ} - g_{mDQ}^2\omega^2}{D_m D_{DD} D_{DQ} - D_{DQ}g_{mDD}^2\omega^2 - D_{DD}g_{mDQ}^2\omega^2 - D_m g_{DDDQ}^2\omega^2}\right|^2, \quad (5)$$

$$|x_{DQ}(\omega)|^2 = \left|\frac{\omega(-iD_m g_{DDDQ} + g_{mDD} g_{mDQ}\omega)}{D_m D_{DD} D_{DQ} - D_{DQ}g_{mDD}^2\omega^2 - D_{DD}g_{mDQ}^2\omega^2 - D_m g_{DDDQ}^2\omega^2}\right|^2. \quad (6)$$

where $D_i(\omega) = \omega_i^2 - \omega^2 - i\gamma_i\omega$ for $i$ = m, DD, DQ. In the quasi-static limit, $\sigma_{sca}(\omega)$ is $(8\pi/3)k^4|\alpha|^2$, where $k = \omega n/c$ is the wave vector of light ($n$, $c$, and $\alpha = Px_{DD}$ are the refractive index of the medium, velocity of light, and polarizability, respectively).[41] By substituting $x_{DD}(t)$ for $\alpha$, one can obtain the cross section:

$$\sigma_{sca}(\omega) \propto \omega^4 |x_{DD}(\omega)|^2. \quad (7)$$

We evaluated the spectral features of $\sigma_{sca}(\omega)$ summarized in Eqs. (1)–(4) using Eq. (7).

We then derived $F_R(\omega)$ using the COM. $F_R(\omega)$ was treated as a de-excitation polarization that induced spontaneous emission from the coupled oscillator. The COM expressed by Eqs. (1)–(3) describe $\sigma_{sca}(\omega)$ as the optical response triggered by light incident from free space, as shown in Eq. (7). Spontaneous emission originates from the transition from an electronic excited state ($S_1$) to ground state ($S_0$) of the excitons. This transition is triggered by the surrounding EM vacuum fluctuations inside the HS. Thus, the driving force of the coupled oscillator is the de-excitation of the molecular exciton.[11,36]

This concept has been modeled using the Heisenberg–Langevin equation, which describes the dynamics of the coupled oscillator initially in its excited state.[42,43] Thus, the driving force term should be the exciton oscillator, as shown in Eq. (1). The dynamics of this type of coupled oscillator have been discussed using the Wigner–Weiskopf approximation.[42,43] The relevance of this approximation was confirmed by evaluating the spontaneous emission spectra of strong-coupling systems.[2,11,36] Thus, we modified the equations of motion for the coupled oscillator as follows:

$$\frac{\partial^2 x_m(t)}{\partial t^2} + \gamma_m \frac{\partial x_m(t)}{\partial t} + \omega_m^2 x_m(t) - g_{mDD}\frac{\partial x_{DD}(t)}{\partial t} - g_{mDQ}\frac{\partial x_{DQ}(t)}{\partial t} = \xi_m(t), \qquad (8)$$

$$\frac{\partial^2 x_{DD}(t)}{\partial t^2} + \gamma_{DD}\frac{\partial x_{DD}(t)}{\partial t} + \omega_{DD}^2 x_{DD}(t) + g_{mDD}\frac{\partial x_m(t)}{\partial t} + g_{DDDQ}\frac{\partial x_{DQ}(t)}{\partial t} = 0, \qquad (9)$$

$$\frac{\partial^2 x_{DQ}(t)}{\partial t^2} + \gamma_{DQ}\frac{\partial x_{DQ}(t)}{\partial t} + \omega_{DQ}^2 x_{DQ}(t) + g_{mDQ}\frac{\partial x_m(t)}{\partial t} + g_{DDDQ}\frac{\partial x_{DD}(t)}{\partial t} = 0. \qquad (10)$$

where $\xi_m(t)$ is the surrounding EM vacuum, which causes fluctuations in the exciton oscillation. The spectra of $x_m$, $x_{DD}$, and $x_{DQ}$ are calculated as the Fourier transform of the autocorrelation function (Wiener–Khinchin theorem) as $F_i(\omega) = \int_0^\infty d\tau e^{-i\omega\tau} \int_0^\infty dt x_i(\tau+t) x_i^*(t)$ for $i$ = m, DD, DQ.[2] If $x_i(\omega) = \int_0^\infty dt x_i(t) e^{-i\omega t}dt$, $F_i(\omega) = |x_i(\omega)|^2$ for the excited state of the coupled system. Therefore, $F_m(\omega)$, $F_{DD}(\omega)$, and $F_{DQ}(\omega)$ can be derived from Eqs. (8)–(10) as follows:

$$F_m(\omega) = \left|\frac{D_{DD}D_{DQ} - g_{DDDQ}^2\omega^2}{D_m D_{DD}D_{DQ} - D_{DQ}g_{mDD}^2\omega^2 - D_{DD}g_{mDQ}^2\omega^2 - D_m g_{DDDQ}^2\omega^2}\right|^2, \qquad (11)$$

$$F_{\text{DD}}(\omega) = \left|\frac{i(D_{\text{DQ}}g_{\text{mDD}}\omega - ig_{\text{mDQ}}g_{\text{DDDQ}}\omega^2)}{D_m D_{\text{DD}} D_{\text{DQ}} - D_{\text{DQ}}g_{\text{mDD}}^2\omega^2 - D_{\text{DD}}g_{\text{mDQ}}^2\omega^2 - D_m g_{\text{DDDQ}}^2\omega^2}\right|^2, \quad (12)$$

$$F_{\text{DQ}}(\omega) = \left|\frac{\omega(iD_{\text{DD}}g_{\text{mDQ}} + g_{\text{mDD}}g_{\text{DDDQ}}\omega)}{D_m D_{\text{DD}} D_{\text{DQ}} - D_{\text{DQ}}g_{\text{mDD}}^2\omega^2 - D_{\text{DD}}g_{\text{mDQ}}^2\omega^2 - D_m g_{\text{DDDQ}}^2\omega^2}\right|^2. \quad (13)$$

In a previous study using Type I dimers, $F_R(\omega)$ spectra were reasonably reproduced by DD-coupled plasmons without considering the emission directly from molecular exciton.[21] Thus, we evaluated the present spectral features of $F_R(\omega)$ summarized in (1)–(4) by $F_{\text{DD}}(\omega)$ and $F_{\text{DQ}}(\omega)$ without considering $F_m(\omega)$. $F_m(\omega)$, $F_{\text{DD}}(\omega)$, and $F_{\text{DQ}}(\omega)$ are intensities of de-excitation polarizations, and do not directly indicate spontaneous emission intensities, because they do not include incoherent processes such as population decay by their energy transfers from the higher energy coupled mode to the lower energy coupled modes.[44–47] Furthermore, they also do not include the quenching processes of higher-order plasmons, which are quite sensitive to the distance between a molecule and the metal surface inside a HS.[21,33] The lack of these incoherent processes in the COM results in the overestimation of the spontaneous emission intensity of $F_m(\omega)$. Our previous success in the reproduction of $F_R(\omega)$ without $F_m(\omega)$ may indicate the existence of this overestimation.[21] However, there is a report suggesting the importance of $F_m(\omega)$.[11] Therefore, our following discussion on the spectral changes in $F_{\text{DD}}(\omega)$ and $F_{\text{DQ}}(\omega)$ is qualitative and not quantitative.

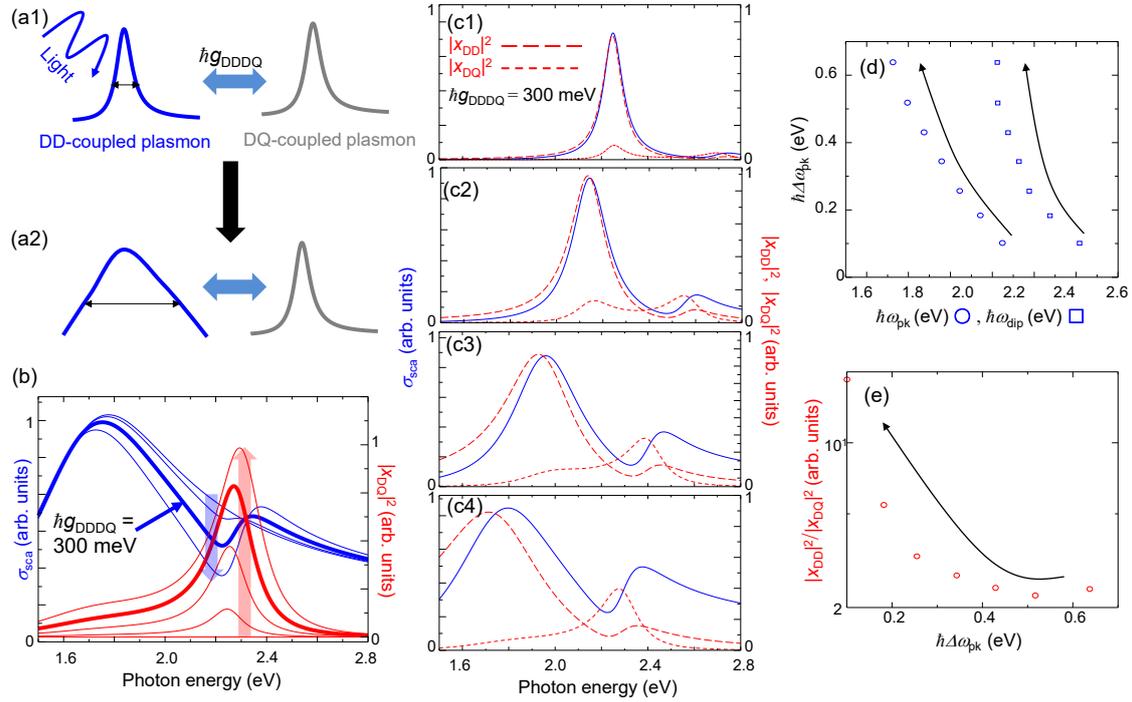

FIG. 6. Schematics of EM coupling between DD- and DQ-coupled plasmon by $\hbar g_{DDDQ}$. DD-coupled plasmon with narrow (a1) and broad (a2) $\hbar\gamma_{DD}$, respectively. (b) Spectra of $\sigma_{sca}(\omega)$ (blue curves) and $|x_{DQ}(\omega)|^2$ (red curves) for various $\hbar g_{DDDQ}$ 0, 100, 200, 300, and 400 meV. The directions of changes from 0 to 400 meV for $\sigma_{sca}(\omega)$ and $|x_{DQ}|^2$ are indicated by blue and red arrows, respectively. The bold curve of $\hbar g_{DDDQ}$ at 300 meV well reproduces the FDTD calculation result in Fig. 4(c4). Spectra of $\sigma_{sca}(\omega)$ (blue curves), $|x_{DD}(\omega)|^2$ (red dashed curves), and $|x_{DQ}(\omega)|^2$ (red dotted curves) by changing $\hbar\omega_{DD}$, $\hbar\omega_{DQ}$, $\hbar\gamma_{DD}$, and $\hbar\gamma_{DQ}$ (eV) as (c1) 2.3, 2.65, 0.1, and 0.15; (c2) 2.2, 2.5, 0.2, and 0.15; (c3) 2.0, 2.35, 0.4, and 0.15; and (c4) 1.9, 2.3, 0.5, and 0.15, respectively. (d) Relationship between $\hbar\Delta\omega_{pk}$ and $\hbar\omega_{pk}$ (blue circles) or $\hbar\omega_{dip}$ (blue squares) for spectra by changing ($\hbar\omega_{DD}$, $\hbar\omega_{DQ}$, $\hbar\gamma_{DD}$, $\hbar\gamma_{DQ}$) eV as (2.3, 2.65, 0.1, 0.15), (2.2, 2.5, 0.2, 0.15), (2.1, 2.4, 0.3, 0.15), (2.0, 2.35, 0.4, 0.15), (1.9, 2.3, 0.5, 0.15), (1.8, 2.25, 0.6, 0.15), and (1.7, 2.25, 0.7, 0.15), respectively. The directions of changes are indicated by the arrows. (e) Relationship between $\hbar\Delta\omega_{pk}$ and $|x_{DD}(\omega_{pk})|^2/|x_{DQ}(\omega_{dip})|^2$ by changing $\hbar\omega_{DD}$, $\hbar\omega_{DQ}$, $\hbar\gamma_{DD}$, and $\hbar\gamma_{DQ}$ (eV) similar to those in (d). The direction of change is indicated by the arrow.

We estimated the EM coupling energy between the DD- and DQ-coupled plasmons using $\sigma_{sca}(\omega)$ spectra shown in Fig. 4, before adding molecular excitons to the COM. In Figs. 4(c1)–4(c4), this EM coupling appears as the spectral dip in $\sigma_{sca}(\omega)$ as $\hbar\Delta\omega_{pk}$ broadens with larger $D_1$. Therefore, we reproduced these spectral dips with increasing $\hbar\Delta\omega_{pk}$, as illustrated in Figs. 6(a1) and 6(a2). First, we determined $g_{DDDQ}$ to reproduce the dip depth of $\sigma_{sca}(\omega)$ in Fig. 4(c4). Figure 6(b) shows $g_{DDDQ}$ dependence of the spectral dips in $\sigma_{sca}(\omega)$. This dip deepened with increasing $g_{DDDQ}$ as indicated by the blue arrow. The excitation polarization intensity of the DQ-coupled plasmon $\propto |x_{DQ}|^2$ increases with increasing $g_{DDDQ}$ (red arrow), indicating that the coherent energy transfer from $|x_{DD}|^2$ to $|x_{DQ}|^2$ determines the dip depth. The dip depth shown in Fig. 4(c4) was reproduced well using a $g_{DDDQ}$ of 300 meV. Therefore, this value was used for $g_{DDDQ}$ in all subsequent calculations. Then, the spectra of $\sigma_{sca}(\omega)$, $|x_{DD}|^2$, and $|x_{DQ}|^2$ were calculated using the values of $\hbar\omega_{DD}$, $\hbar\omega_{DQ}$, and $\hbar\gamma_{DD}$ from $\hbar\omega_{pk}$, $\hbar\omega_{dip}$, and $\hbar\Delta\omega_{pk}$ of the FDTD calculation in Figs. 4(c1)–4(c4). The value of $\hbar\gamma_{DQ}$ is fixed to be 150 meV considering the rather constant linewidths of $F_R(\omega)$ in Figs. 4(c1)–4(c4) compared with those of $\sigma_{sca}(\omega)$. The constant linewidths mean that the radiative loss of DQ-coupled plasmon is significantly small compared with DD-coupled plasmon.[24] Figures 6(c1)–6(c4) show that the $\sigma_{sca}(\omega)$ spectra by the COM well reproduce those by the FDTD calculation in Figs. 4(c1)–4(c4). For the narrow $\hbar\Delta\omega_{pk}$ as in Fig. 6(c1), $\sigma_{sca}(\omega)$ is almost identical to $|x_{DD}|^2$, and the dip appears negligible. With increasing $\hbar\Delta\omega_{pk}$, the dip appears near the peak position of $|x_{DQ}|^2$, indicating the energy transfer from $|x_{DD}|^2$ to $|x_{DQ}|^2$ via $g_{DDDQ}$ with spectral overlapping between $|x_{DD}|^2$ and $|x_{DQ}|^2$. When the dip appears in $\sigma_{sca}(\omega)$, the intensity of $|x_{DQ}|^2$ becomes comparable to that of $|x_{DD}|^2$. The $F_R(\omega_{dip}) \gg F_R(\omega_{pk})$ satisfy for large $\hbar\Delta\omega_{pk}$ in the FDTD calculation [Fig. 4(c4)]; however, $|x_{DD}|^2$ is larger than the $|x_{DQ}|^2$ even for large $\hbar\Delta\omega_{pk}$ [Fig. 6(c4)]. This contradiction is because the $F_R(\omega)$ calculated by the FDTD method indicates the localized intensity only inside the HS, whereas $|x_{DD}|^2$ means the whole intensity of the DD-coupled polarization, including outside the HS. Figure 6(d) shows $\hbar\omega_{pk}$ and $\hbar\omega_{dip}$ dependences of $\hbar\Delta\omega_{pk}$, reproducing the results of FDTD calculation in Fig. 4(e). Figure 6(e) shows $\hbar\Delta\omega_{pk}$ dependence of the ratio $|x_{DD}|^2/|x_{DQ}|^2$, also qualitatively reproducing the results of FDTD calculation in Fig. 4(f). These reproductions of results indicate that the DQ-coupled resonance dominates EM enhancement of dimers having the large $\hbar\Delta\omega_{pk}$.

Further, the EM coupling among the DD-, DQ-coupled plasmons, and molecular exciton were evaluated using the COM by broadening the linewidth of the DD-coupled resonance, as illustrated in Figs. 7(a1) and 7(a2). The values of $\hbar\omega_{DD}$, $\hbar\omega_{DQ}$, $\hbar\gamma_{DD}$, and $\hbar\gamma_{DQ}$ are the same as those used in Figs. 6(c1)–6(c4). The values of $\hbar g_{mDD}$ and $\hbar g_{mDQ}$ are set to be identical because both share the common HS with identical mode volume and the number of dye molecules.[21] The $g_{DDDQ}$ is fixed at 300 meV based on the analysis of dip depth of $\sigma_{sca}(\omega)$, as discussed using Figs. 4 and 6. The values of $\hbar g_{mDD}$ of silver NP dimers have been evaluated to be approximately 200 meV under near-single molecule SERRS condition using the spectral changes in $\sigma_{sca}(\omega)$ during SERRS quenching.[10,21,22]

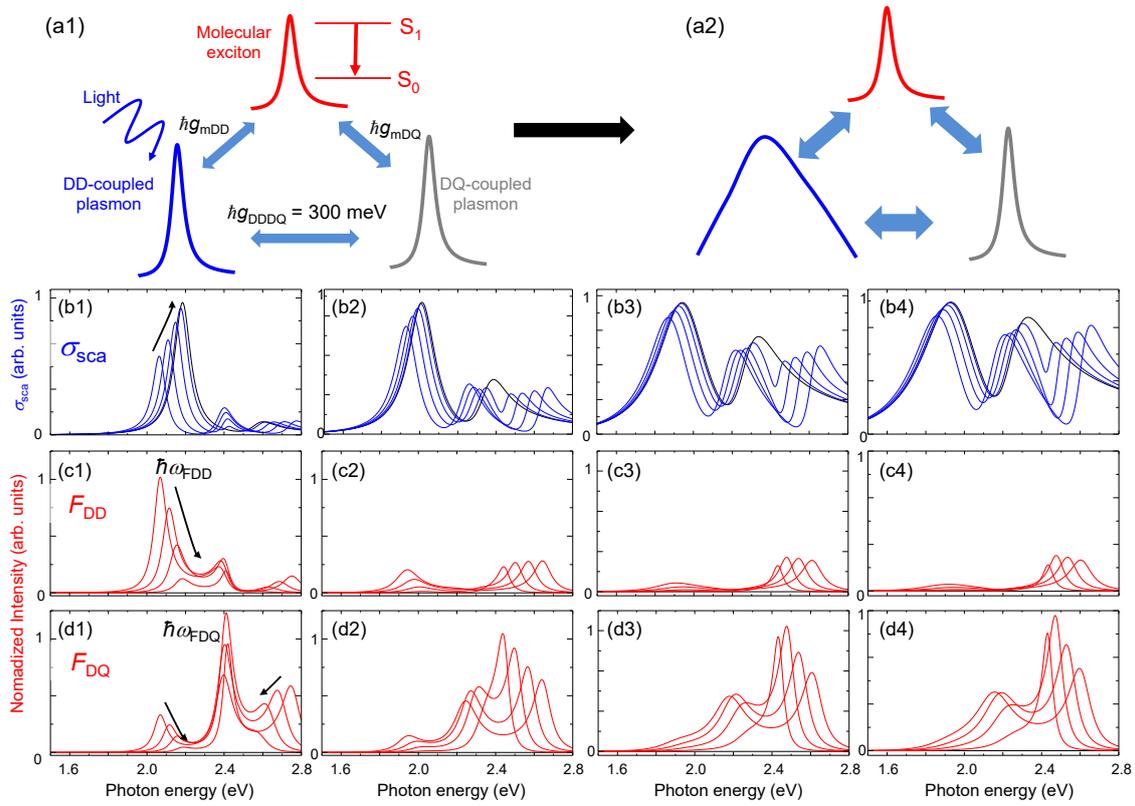

FIG. 7. Schematics of EM coupling between DD-coupled plasmon, DQ-coupled plasmon, and molecular exciton with coupling energy of $\hbar g_{mDD}$ (= $\hbar g_{mDQ}$) with narrow (a1) and broad (a2) $\hbar \gamma_{DD}$. Rayleigh scattering is triggered by light indicated with a blue wavy arrow. Ultrafast SEF is triggered by transition ($S_1$ to $S_0$) indicated with the red energy-level diagram. Spectra of $\sigma_{sca}(\omega)$, $F_{DD}(\omega)$, and $F_{DQ}(\omega)$ by changing ($\hbar\omega_{DD}, \hbar\omega_{DQ}, \hbar\gamma_{DD}, \hbar\gamma_{DQ}$) eV: (b1–d1) (2.3, 2.65, 0.1, 0.15; (b2–d2) (2.2, 2.5, 0.2, 0.15); (b3–d3) (2.0, 2.35, 0.4, 0.15) ; and (b4–d4) (1.9, 2.3, 0.5, 0.15), respectively. $\hbar g_{mDD}$: 400 (blue curves), 300 (blue curves), 200 (blue curves), 100 (blue curves), 0 (black curves) meV. $F_{DD}(\omega)$ and $F_{DQ}(\omega)$ are normalized by their largest peak intensities for each ($\hbar\omega_{DD}, \hbar\omega_{DQ}, \hbar\gamma_{DD}, \hbar\gamma_{DQ}$). The directions of spectral changes are indicated by black arrows.

The value of $\hbar g_{mDD}$ was also confirmed by Rabi splitting in $\sigma_{sca}(\omega)$ spectra.[7–9,48] Under the present experimental condition, the number of molecules in a HS may be more than one. Indeed, we reasonably reproduced the experimentally obtained spectral changes in $\sigma_{sca}(\omega)$ by setting the $\hbar g_{mDD}$ from 100 to 600 meV.[10,21,22] Therefore, the maximum value for $\hbar g_{mDD}$ (= $\hbar g_{mDQ}$) was set to 400 meV. The experimental spectral changes in Figs. 2 and 3 were evaluated by decreasing $\hbar g_{mDD}$ from 400 to 0 meV. Figures 7(b1)–7(b4), 7(c1)–7(c4), and 7(d1)–7(d4) exhibit the spectral changes in $\sigma_{sca}(\omega)$, $F_{DD}(\omega)$, and $F_{DQ}(\omega)$ with decreasing $\hbar g_{mDD}$. The blue-shifts in $\hbar \omega_{pk}$ are consistent with the experimentally observed shifts. For the coupled oscillator with the narrow linewidth of $\hbar \gamma_{DD}$, as shown in Figs. 7(b1), 7(c1), and 7(d1), the spectral peak changes in $\sigma_{sca}(\omega)$ are similar to those of $F_{DD}(\omega)$. These changes reproduce the properties of the Type I dimer, indicating that $F_{DD}(\omega)$ mainly generates ultrafast SEF for Type I dimers. For the coupled oscillator with the broad linewidth of $\hbar \gamma_{DD}$, as shown in Figs. 7(b4), 7(c4), and 7(d4), the spectral dip changes in $\sigma_{sca}(\omega)$ are similar to the spectral peak changes in $F_{DQ}(\omega)$. These changes reproduce the properties of Type II dimers, indicating that $F_{DQ}(\omega)$ mainly generates ultrafast SEF. In Figs. 7(c1)–7(c4) and 7(d1)–7(d4), the contribution of molecular exciton around 2.5 eV appears to be always overestimated. This overestimation may be due to the lack of incoherent population decay from the molecular exciton to the DD- or DQ-coupled plasmons in the COM.[44–47] Indeed, $F_m(\omega)$, $F_{DD}(\omega)$, and $F_{DQ}(\omega)$ are de-excitation polarizations that do not include population decay, which results in spontaneous emission from the lower-energy coupled resonances. Such a population decay process is discussed to explain the spectral shapes in surface-enhanced spontaneous emission.[36,44–47]

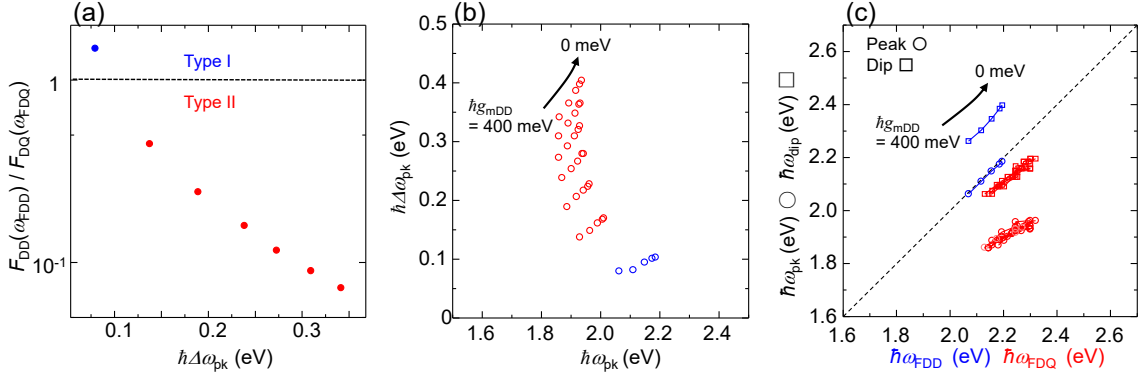

FIG. 8. (a) Relationship between $\hbar\Delta\omega_{pk}$ and $F_{DD}(\omega_{FDD})/F_{DQ}(\omega_{FDQ})$. (b) Relationship between $\hbar\Delta\omega_{pk}$ and $\hbar\omega_{pk}$. (c) Relationship between $\hbar\omega_{FDD}$ and $\hbar\omega_{pk}$ (blue circles) or $\hbar\omega_{dip}$ (blue squares). Relationship between $\hbar\omega_{FDQ}$ and $\hbar\omega_{pk}$ (red circles) or $\hbar\omega_{dip}$ (red squares). In all figures, the blue and red symbols correspond to Type I and II dimers, respectively.

We examined the relationships between $\hbar\omega_{pk}$, $\hbar\Delta\omega_{pk}$, and $\hbar\omega_F$ as summarized in Fig. 3 using the calculation results in Fig. 7. Figure 8(a) shows the ratio $F_{DD}(\omega_{FDD})/F_{DQ}(\omega_{FDQ})$ with respect to $\hbar\Delta\omega_{pk}$, where $\hbar\omega_{FDD}$ and $\hbar\omega_{FDQ}$ are defined as the lowest and the second lowest peak energies directly related to DD- or DQ-coupled plasmon, respectively, as indicated in Figs. 7(c1) and 7(d1). The ratio sharply decreases with increasing $\hbar\Delta\omega_{pk}$, supporting that the DQ-coupled resonance predominantly generates EM enhancement due to the efficient energy transfer from DD- to DQ-coupled plasmon with their spectral overlapping. This result is consistent with the FDTD calculation results shown in Fig. 4(f).

Figure 8(b) shows the $\hbar g_{mDD}$ (= $\hbar g_{mDQ}$) dependence of the relationships between $\hbar \omega_{pk}$ and $\hbar \Delta \omega_{pk}$. The $\hbar \omega_{pk}$ and $\hbar \Delta \omega_{pk}$ exhibit blue shifts and broadening, respectively, with deceasing $\hbar g_{mDD}$ from 400 to 0 meV. The amount of blue shift in $\hbar \omega_{pk}$ appears to decrease with decreasing $\hbar \omega_{pk}$, supporting the experimentally observed tendency in Fig. 3(a). This tendency is because of the inefficient EM coupling between DD-coupled resonance and exciton resonance as the spectral separation between $\hbar \omega_{DD}$ and $\hbar \omega_m$ increases.[10] Figure 8(c) exhibits the relationships between $\hbar \omega_{FDD}$ (or $\hbar \omega_{FDQ}$) and $\hbar \omega_{pk}$ (or $\hbar \omega_{dip}$) with $\hbar g_{mDD}$ (= $\hbar g_{mDQ}$). The blue represents the relationship under the condition $F_{DD}(\omega_{FDD}) > F_{DQ}(\omega_{FDQ})$ and the red represents the relationship under the condition $F_{DD}(\omega_{FDD}) < F_{DQ}(\omega_{FDQ})$, separated by the dashed line in Fig. 8(a). These relationships qualitatively reproduce the experimental results in Figs. 3(c) and 3(d), in which the dimers with narrow and broad linewidths exhibit $\hbar \omega_{pk} \sim \hbar \omega_{FDD}$ and $\hbar \omega_{dip} \sim \hbar \omega_{FDQ}$, respectively. Thus, the relationships in Fig. 8(c) indicate that the emission of the ultrafast SEF from the DQ-coupled resonance becomes dominant via coherent energy transfer from the DD- to DQ-coupled resonance, with a spectral overlap between the DQ- and DD-coupled resonances.

We have discussed the experimental spectral changes in $\sigma_{sca}(\omega)$ and $F_R(\omega)$ with respect to $\hbar g_{mDD}$ (= $\hbar g_{mDQ}$) in the COM using Figs. 7 and 8. The spectral changes of Type I and II were dominated by DD- and DQ-coupled resonances, respectively. To visualize more clearly the spectral features of Type I and II by the COM, we converted the $\hbar g_{\mathrm{mDD}}$ dependence of $\sigma_{\mathrm{sca}}(\omega)$, $F_{\mathrm{DD}}(\omega)$, and $F_{\mathrm{DQ}}(\omega)$ in Figs. 7(b1), 7(c1), and 7(d1) and Figs. 7(b4), 7(c4), and 7(d4) into the contour maps, as shown in Fig. 9. The calculated changes in $\hbar\omega_{\mathrm{pk}}$ (or $\hbar\omega_{\mathrm{dip}}$) and $\hbar\Delta\omega_{\mathrm{pk}}$ of $\sigma_{\mathrm{sca}}(\omega)$ in Fig. 9(a1) and Fig. 9(b1) well reproduce the experimental results in Figs. 2(d1)–2(f1) and Figs. 2(j1)–2(i1) for the Type I and II, respectively. Then, we discuss $F_{\mathrm{R}}(\omega)$ in Figs. 2(d2)–2(f2) and Figs. 2(j2)–2(i2) for Type I and II, respectively, using $F_{\mathrm{DD}}(\omega)$ and $F_{\mathrm{DQ}}(\omega)$. Figure 9(a2) (not 9(a3)) well reproduces $F_{\mathrm{R}}(\omega)$ in Figs. 2(d2)–2(f2), which exhibit the features of Type I, indicating that this $F_{\mathrm{R}}(\omega)$ is mainly generated by the $F_{\mathrm{DD}}(\omega)$. Figure 9(b3) (not 9(b2)) well reproduce $F_{\mathrm{R}}(\omega)$ in Figs.

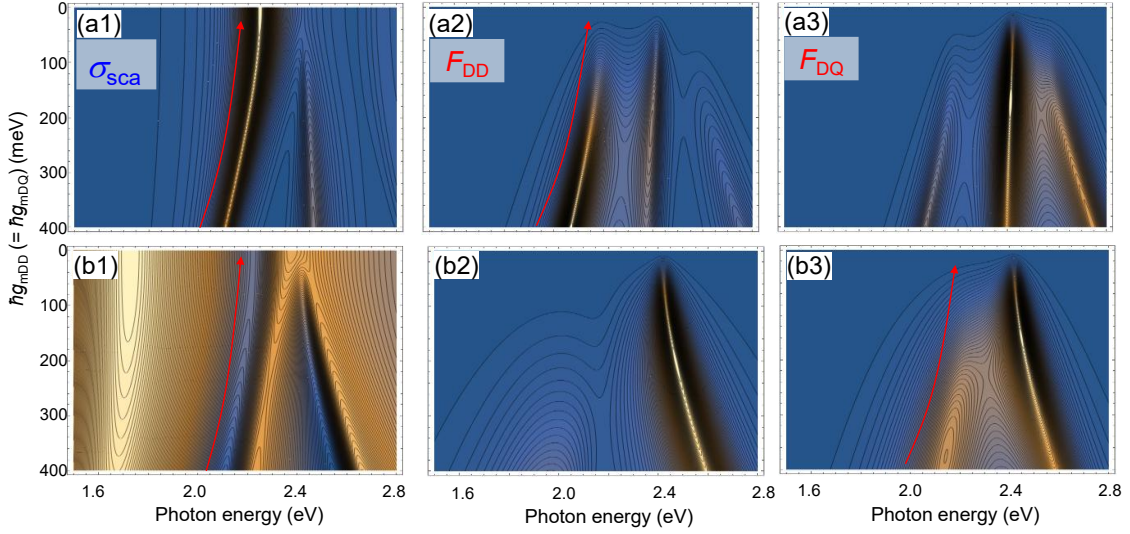

FIG. 9. (a1–a3) $\hbar g_{\mathrm{mDD}}$ (= $\hbar g_{\mathrm{mDQ}}$) dependences of $\sigma_{\mathrm{sca}}(\omega)$, $F_{\mathrm{DD}}(\omega)$, and $F_{\mathrm{DQ}}(\omega)$ expressed as the contour maps. $\hbar\omega_{\mathrm{DD}}$, $\hbar\omega_{\mathrm{DQ}}$, $\hbar\gamma_{\mathrm{DD}}$, and $\hbar\gamma_{\mathrm{DQ}}$ (eV): 2.25, 2.5, 0.1, and 0.15 for Type I dimer. (b1–b3) $\hbar g_{\mathrm{mDD}}$ (= $\hbar g_{\mathrm{mDQ}}$) dependences of $\sigma_{\mathrm{sca}}(\omega)$, $F_{\mathrm{DD}}(\omega)$, and $F_{\mathrm{DQ}}(\omega)$ expressed as the contour maps. $\hbar\omega_{\mathrm{DD}}$, $\hbar\omega_{\mathrm{DQ}}$, $\hbar\gamma_{\mathrm{DD}}$, and $\hbar\gamma_{\mathrm{DQ}}$ (eV): 2.03, 2.19, 0.4, and 0.15 for Type II dimer.

2(j2)–2(i2), indicating that $F_R(\omega)$ is generated by $F_{DQ}(\omega)$. This consistency between the experimental results and calculations demonstrates that the experimentally observed changes in $\sigma_{sca}(\omega)$ and $F_R(\omega)$ are induced by the decrease in both $\hbar g_{mDD}$ and $\hbar g_{mDQ}$. These results also indicate that the spectral properties of Type I and II dimers are improved by the absence or presence of coherent energy transfer from DD- to DQ-coupled resonances, owing to their spectral overlap.

Finally, we briefly discuss other factors related to the observed spectral changes in $\sigma_{sca}(\omega)$ and $F_R(\omega)$ as summarized in (1)–(4). The first factor is anion-induced carving of dimer surfaces. Maruyama et al.[49] reported the changes in $\sigma_{sca}(\omega)$ by the carving due to addition of the solutions containing anions Br$^-$, SCN$^-$, and CN$^-$ at concentrations of 10 mM to silver NP aggregates. However, we did not observe such changes in $\sigma_{sca}(\omega)$ for SERRS-inactive dimers under the present Cl$^-$ concentration of 1.0 mM. Thus, anion-induced carving may play a minor role in the observed spectral changes. The second factor is the photoinduced melting or decomposition of the dimer surfaces. This phenomenon was observed as changes in $\sigma_{sca}(\omega)$ for silver NPs by laser irradiation at a power of 6.2 kW/cm$^2$.[50] However, changes in $\sigma_{sca}(\omega)$ were not observed for SERRS-inactive dimers under the present laser irradiation power of 3.5 W/cm$^2$, indicating that photoinduced melting or decomposition may be within the experimental error level under the present irradiation conditions. Thus, we conclude that the changes in EM coupling play a major role in the observed spectral changes in $\sigma_{sca}(\omega)$ and $F_R(\omega)$.

## IV. CONCLUSION

In this study, we investigated the spectral changes in $F_R(\omega)$ derived using ultrafast SEF of NP dimers having fluorescent molecules within the HSs, with a focus on the EM coupling between the DQ-coupled plasmons of asymmetric dimers. The DQ-coupled resonance of the asymmetric dimers appeared as a peak and dip in the $F_R(\omega)$ and $\sigma_{sca}(\omega)$ spectra, respectively. Both the $F_R(\omega)$ peak and $\sigma_{sca}(\omega)$ dip exhibited blue shift during SERRS quenching. These static and temporal spectral properties were examined using a COM composed of three oscillators representing DD-coupled plasmons, DQ-coupled plasmons, and molecular excitons. The calculation conditions for the COM were determined by FDTD calculations, reproducing the experimental $\sigma_{sca}(\omega)$ properties. These static and temporal spectral properties were reproduced well by the COM with variations in linewidths of the DD-coupled resonances and decreasing coupling energies among the oscillators, respectively. This reproduction indicates that this method combining ultrafast SEF and COM is useful for evaluating the EM coupling between subradiant plasmons and molecular excitons and thus contributes to connecting surface-enhanced spectroscopy and various phenomena related to cavity QED.[18–20,51,52]

## SUPPLEMENTARY MATERIAL

See the supplementary material for more details on an intermediate type between Type I and II.

## ACKNOWLEDGMENTS

This work was supported by a JSPS KAKENHI Grant-in-Aid for Scientific Research (C) (Grant No. 25K08520 and 25K08506).

## DATA AVAILABILITY

The data that support the findings of this study are available from the corresponding authors upon reasonable request.

## REFERENCES

[1]P. Törmä and W. L. Barnes, "Strong coupling between surface plasmon polaritons and emitters: a review," Rep. Prog. Phys. **78**, 013901 (2015).

[2]M. Pelton, S. D. Storm, and H. Leng, "Strong coupling of emitters to single plasmonic nanoparticles: exciton-induced transparency and Rabi splitting," Nanoscale **11**, 14540 (2019).

[3]Y. Tomoshige, M. Tamura, T. Yokoyama, and H. Ishihara, "Coherent coupling among plasmons, electron– hole pairs, and light: energy transparency, imaging, and efficient hot-carrier generation," Nanophotonics **14**, 1157–1169 (2025).

[4]R. Esteban, J. J. Baumberg, and J. Aizpurua, "Molecular optomechanics approach to surface-enhanced Raman scattering," Acc. Chem. Res. **55**, 1889–1899 (2022).

[5]A. Mandal, M. A. D. Taylor, B. M. Weight, E. R. Koessler, X. Li, and P. Huo, "Theoretical advances in polariton chemistry and molecular cavity quantum electrodynamics," Chem. Rev. **123**, 9786−9879 (2023).

[6]J. Yuen-Zhou, Wei Xiong and T. Shegai, "Polariton chemistry: Molecules in cavities and plasmonic media," J. Chem. Phys. **156**, 030401 (2022).

[7]G. Zengin, M. Wersäll, S. Nilsson, T. J. Antosiewicz, M. Käll, and T. Shegai, "Realizing strong light-matter interactions between single-nanoparticle plasmons and molecular excitons at ambient conditions," Phys. Rev. Lett. **114**, 157401 (2014).

[8]A. E. Schlather, N. Large, A. S. Urban, P. Nordlander, and N. J. Halas, "Near-field mediated plexcitonic coupling and giant Rabi splitting in individual metallic dimers," Nano Lett. **13**, 3281 (2013).

[9]F. Nagasawa, M. Takase, and K. Murakoshi, "Raman enhancement via polariton states produced by strong coupling between a localized surface plasmon and dye excitons at metal nanogaps," J. Phys. Chem. Lett. **5**, 14 (2014).

[10]T. Itoh, Y. S. Yamamoto, H. Tamaru, V. Biju, S. Wakida, and Y. Ozaki, "Single-molecular surface-enhanced resonance Raman scattering as a quantitative probe of local electromagnetic field: The case of strong coupling between plasmonic and excitonic resonance," Phys. Rev. B **89**, 195436 (2014).

[11]H. Leng, B. Szychowski, M.-C. Daniel, and M. Pelton, "Strong coupling and induced transparency at room temperature with single quantum dots and gap plasmons," Nat. Comm. **9**, 4012 (2018).

[12]S. Nie and S. Emory, "Probing single molecules and single nanoparticles by surface-enhanced Raman scattering," Science **275**, 1102 (1997).

[13]K. Kneipp, Y. Wang, H. Kneipp, L. Perelman, I. Itzkan, R. R. Dasari, and M. Feld, "Single molecule detection using surface-enhanced Raman scattering (SERS)," Phys. Rev. Lett.. **78**, 1667 (1997).

[14]C. Ciraci, R. T. Hill, J. J. Mock, Y. Urzhumov, A. I. Fernandez-Dominguez, S. A. Maier, J. B. Pendry, A. Chilkoti, and D. R. Smith, "Probing the ultimate limits of plasmonic enhancement," Science **337**, 1072–1074 (2012).

[15]W. Zhu and K. B. Crozier, "Quantum mechanical limit to plasmonic enhancement as observed by surface-enhanced Raman scattering," Nat. Commun. **5**, 5228 (2014).

[16]T. Itoh and Y. S. Yamamoto, "Plasmon-enhanced two photon excited emission from edges of one-dimensional plasmonic hotspots with continuous-wave laser excitation," J. Chem. Phys. **161**, 164704 (2024).

[17]T. Itoh and Y. S. Yamamoto, "Ultrabroad resonance of localized plasmon on a nanoparticle coupled with surface plasmon on a nanowire enabling two-photon excited emission via continuous-wave laser," J. Chem. Phys. **163**, 034711 (2025).

[18]T. Itoh, Y. S. Yamamoto, and Y. Ozaki, "Plasmon-enhanced spectroscopy of absorption and spontaneous emissions explained using cavity quantum optics," Chem. Soc. Rev. **46**, 3904 (2017).

[19]T. Itoh and Y. S. Yamamoto, "Between plasmonics and surface-enhanced resonant Raman spectroscopy: toward single-molecule strong coupling at a hotspot," Nanoscale **13**, 1566 (2021).

[20]T. Itoh, M. Prochazka, Z.-C. Dong, W. Ji, Y. S. Yamamoto, Y. Zhang, and Y. Ozaki, "Toward a new era of SERS and TERS at the nanometer scale: from fundamentals to innovative applications," Chem. Rev. **123**, 1552 (2023).

[21]T. Itoh and Y. S. Yamamoto, "Reproduction of surface-enhanced resonant Raman scattering, and fluorescence spectra of a strong coupling system composed of a single silver nanoparticle dimer and a few dye molecules," J. Chem. Phys. **149**, 244701 (2018); T. Itoh, Y. S. Yamamoto, and T. Okamoto, "Absorption cross-section spectroscopy of a single strong-coupling system between plasmon and molecular exciton resonance using a single silver nanoparticle dimer generating surface-enhanced resonant Raman scattering," Phys. Rev. B **99**, 235409 (2019).

[22]T. Itoh, Y. S. Yamamoto, and T. Okamoto, "Anti-crossing property of strong coupling system of silver nanoparticle dimers coated with thin dye molecular films analyzed by electromagnetism," J. Chem. Phys. **152**, 054710 (2020).

[23]L. V. Brown, H. Sobhani, J. B. Lassiter, P. Nordlander, and N. J. Halas, "Heterodimers: plasmonic properties of mismatched nanoparticle pairs," ACS Nano **4**, 819–832 (2010).

[24]T. Itoh and Y. S. Yamamoto, "Contribution of subradiant plasmon resonance to electromagnetic enhancement in resonant Raman with fluorescence examined by single silver nanoparticle dimers," J. Phys. Chem. C **127**, 5886–5897 (2023).

[25]T. Itoh and Y. S. Yamamoto, "Correlated polarization dependences between surface-enhanced resonant Raman scattering and plasmon resonance elastic scattering showing spectral uncorrelation to each other," J. Phys. Chem. B **127**, 4666–4675 (2023).

[26]T. Itoh and Y. S. Yamamoto, "Demonstration of electromagnetic enhancement correlated to optical absorption of single plasmonic system coupled with molecular excitons using ultrafast surface-enhanced fluorescence," J. Chem. Phys. **159**, 2 034709 (2023).

[27]T. Itoh and Y. S. Yamamoto, "Electromagnetic enhancement spectra of one-dimensional plasmonic hotspots along silver nanowire dimer derived via surface-enhanced fluorescence," J. Chem. Phys. **160**, 024703 (2024).

[28]T. Itoh, Y. S. Yamamoto, H. Tamaru, V. Biju, N. Murase, and Y. Ozaki, "Excitation laser energy dependence of surface-enhanced fluorescence showing plasmon-induced ultrafast electronic dynamics in dye molecules," Phys. Rev. B **87**, 235408 (2013).

[29]Y. S. Yamamoto and T. Itoh, "Proving insight into subradiant plasmon resonance in surface-enhanced Raman spectroscopy: a short review," Nano Futures **10**, 012001 (2026).

[30]P. Lee and D. Meisel, "Adsorption and surface-enhanced Raman of dyes on silver and gold sols," J. Phys. Chem. **86**, 3391 (1982).

[31]E. C. Le Ru, M. Meyer, and P. G. Etchegoin, "Proof of single-molecule sensitivity

in surface enhanced Raman scattering (SERS) by means of a two-analyte technique," J. Phys. Chem. B **110**, 1944 (2006).

[32]A. B. Zrimsek, A. I. Henry, and R. P. Van Duyne, "Single molecule surface-enhanced Raman spectroscopy without nanogaps," J. Phys. Chem. Lett. **4**, 3206 (2013).

[33]T. Itoh, M. Iga, H. Tamaru, K. Yoshida, V. Biju, and M. Ishikawa, "Quantitative evaluation of blinking in surface enhanced resonance Raman scattering and fluorescence by electromagnetic mechanism," J. Chem. Phys. **136**, 024703 (2012).

[34]B. Rousseaux, D. G. Baranov, T. J. Antosiewicz, T. Shegai, and G. Johansson, "Strong coupling as an interplay of quantum emitter hybridization with plasmonic dark and bright modes," Phys. Rev. R **2**, 033056 (2020).

[35]J. Qin, Y. -H. Chen, Z. Zhang, Y. Zhang, R. J. Blaikie, B. Ding, and M. Qiu, "Revealing strong plasmon-exciton coupling between nanogap resonators and two-dimensional Semiconductors at ambient conditions," Phys. Rev. Lett. **124**, 063902 (2020).

[36]H. Hu, Z. Shi, S. Zhang, and H. Xu, "Unified treatment of scattering, absorption, and luminescence spectra from a plasmon–exciton hybrid by temporal coupled-mode theory," J. Chem. Phys. **155**, 074104 (2021).

[37]H. Kuwata, H. Tamaru, K. Esumi, and K. Miyano, "Resonant light scattering from metal nanoparticles: Practical analysis beyond Rayleigh approximation," Appl. Phys. Lett. **83**, 4625 (2003).

[38]P. B. Johnson and R. W. Christy, "Optical constants of the noble metals," Phys. Rev. B 6, 4370–4379 (1972).

[39]J. Ye, F. Wen, H. Sobhani, J. B. Lassiter, P. V. Dorpe, P. Nordlander, and N. J. Halas, "Plasmonic nanoclusters: near field properties of the Fano resonance interrogated with SERS," Nano Lett. **12**, 1660−1667 (2012).

[40]W. Liu and Y. S. Kivshar, "Generalized Kerker effects in nanophotonics and meta-optics," Opt. Express **26**, 13085−13105 (2018).

[41]C. F. Bohren and D. R. Huffman, *Absorption and Scattering of Light by Small Particles* (Wiley, New York, 1983).

[42]G. Cui and M. G. Raymer, "Emission spectra and quantum efficiency of single-photon sources in the cavity-QED strong-coupling regime," Phys. Rev. A **73**, 053807 (2006).

[43]F. P. Laussy, E. del Valle, and C. Tejedor, "Luminescence spectra of quantum dots in microcavities. I. Bosons," Phys. Rev. B **79**, 235325 (2009).

[44]Y. Wei, Z. Liao, and X.-h. Wang, "Emission spectrum and photon statistics in cavity-QED system under incoherent pumping," Phys. Lett. A **526**, 129965 (2024).

[45]F. Tassone, C. Piermarocchi, V. Savona, A. Quattropani, and P. Schwendimann, "Bottleneck effects in the relaxation and photoluminescence of microcavity polaritons," Phys. Rev. B **56**, 7554 (1997).

[46]D. M. Coles, N. Somaschi, P. Michetti, C. Clark, P. G. Lagoudakis, P. G. Savvidis, and D. G. Lidzey, "Polariton-mediated energy transfer between organic dyes in a strongly coupled optical microcavity," Nat. Mat. **13**, 712-719 (2014).

[47]Y. Wei, K. Liu, J. Tian, Y. Yang, J. Guo, J.-H. Zhong, and B. Liu, "Photoluminescence Properties of strongly coupled gold plasmonic nanoarray and

molecular aggregates with altered photostability," Adv. Opt. Mater. **14**, e02304 (2026).

[48]R. Chikkaraddy, B. de Nijs, F. Benz, S. J. Barrow, O. A. Scherman, E. Rosta, A. Demetriadou, P. Fox, O. Hess, and J. J. Baumberg, "Single-molecule strong coupling at room temperature in plasmonic nanocavities, Nature **535**, 127 (2016).

[49]Y. Maruyama and M. Futamata, "Anion induced SERS activation and quenching for R6G adsorbed on Ag nanoparticles," Chem. Phys. Lett. **448**, 93 (2007).

[50]Y. Maruyama, M. Ishikawa, and M. Futamata, "Thermal activation of blinking in SERS signal," J. Phys. Chem. B **108**, 673 (2004).

[51]J. Yi et al., "Surface-enhanced Raman spectroscopy: A half-century historical perspective," Chem. Soc. Rev. **54**, 1453–1551 (2025).

[52]J. Langer et al., "Present and future of surface-enhanced Raman scattering," ACS Nano **14**, 28–117 (2020).